\documentclass[12pt]{JHEP3}
\usepackage{amsmath,amsfonts,amssymb,dsfont,ifthen,epsfig,slashed, cancel}

\newcommand{\be}{\begin{equation}}
\newcommand{\ee}{\end{equation}}
\newcommand{\ba}{\begin{eqnarray}}
\newcommand{\ea}{\end{eqnarray}}
\newcommand{\bas}{\begin{eqnarray*}}
\newcommand{\eas}{\end{eqnarray*}}
\newcommand{\bc}{\begin{center}}
\newcommand{\ec}{\end{center}}
\newcommand{\nn}{\nonumber}

\newcommand{\comment}[1]{}

\newboolean{Notes}\setboolean{Notes}{true}
\newcommand{\notes}[1]
            {\ifthenelse{\boolean{Notes}}{{\tt #1}}{}}

\newcommand{\csch}{\mathop{\rm csch\,}}
\newcommand{\sech}{\mathop{\rm sech\,}}

\preprint{arXiv:0711.1298 [hep-th]}

\title{Gauge/Gravity Theory with Running Dilaton and Running Axion}
\author{Girma Hailu\thanks{hailu@lepp.cornell.edu}
\\
Newman Laboratory for Elementary Particle Physics \\
Cornell University\\
Ithaca, NY 14853}

\abstract{

We present a new gauge/gravity duality construction of the Klebanov-Strassler throat which takes corrections to the anomalous mass dimension proposed in \cite{Hailu:2006uj} into account on the gauge theory side and both the dilaton and the axion run on the gravity side. The corresponding supergravity solutions are found using equations for type IIB flows with $\mathcal{N}=1$ supersymmetry obtained in \cite{Hailu:2007ae}.
We find that magnetic couplings of the axion to D7-branes filling 4-d spacetime and wrapping 4-cycles at locations of duality transitions and invisible Dirac 8-branes whose worldvolume emanates from the worldvolume of the D7-branes are the sources for the runnings of the dilaton and the axion.
Our construction provides the first explicit example of a gauge/gravity duality mapping with a running dilaton or a running axion which is an important component towards finding gravity duals to gauge theories with physically more interesting renormalization group flows such as pure confining gauge theories in four dimensions.  The D7-branes also serve as gravitational source for Seiberg duality transitions.
The supergravity background has distinct features which could be useful for constructing cosmological models and studying possibilities for probing stringy signatures from the early universe.
}

\begin{document}

\section{Introduction\label{intr}}

The Klebanov-Strassler throat \cite{Klebanov:2000hb,Klebanov:2000nc,Klebanov:1999rd, Gubser:1998fp,  Klebanov:1998hh} is a setting for the only explicit example of a gauge/gravity duality \cite{Maldacena:1998re,Gubser:1998bc,Witten:1998qj} with  $\mathcal{N}=1$ supersymmetry on  nonsingular supergravity background with both the gauge and the gravity theories explicitly known.
The Maldacena-Nunez solution \cite{Maldacena:2000yy, Chamseddine:1997nm}
is the other nonsingular supergravity background with $\mathcal{N}=1$ supersymmetry.
Numerical supergravity solutions which interpolate between the Klebanov-Strassler and the Maldacena-Nunez solutions were obtained in \cite{Butti:2004pk}.
In the Klebanov-Strassler throat, $\mathcal{N}=1$ supersymmetric $SU(N+M)\times SU(N)$ gauge theory with bifundamental chiral superfields and a quartic tree level superpotential undergoing a cascade of Seiberg duality transitions in four dimensions is dual to type IIB string theory with $N$ D3- and $M$ D5-branes on $AdS_5\times T^{1,1}$ background.
The throat has features which make it physically interesting. First, it has minimal ($\mathcal{N}=1$) supersymmetry and low energy and minimal supersymmetry is, arguably, the most interesting possible extension of the standard model of particle physics, not only does it give a stabilized electroweak scale via cancelation of divergences between bosonic and fermionic loops but it also provides the possibility for a natural unification of the electromagnetic, the weak and the strong interactions into a single more fundamental interaction \cite{Dimopoulos:1981zb}.
Second, the couplings in the gauge theory undergo interesting nonconformal renormalization group flow. Third, the supergravity background has nonsingular geometry which allows well-defined physics at the tip of the throat which is useful for studying possibilities that a universe like ours could reside at the bottom of the throat. Fourth, the throat is warped which gives a hierarchy of scales \cite{Giddings:2001yu} whereby the Planck scale gets exponentially suppressed and also leads to flattening of potentials in brane inflation \cite{Dvali:1998pa,Kachru:2003sx}.  The supergravity background has been used for constructing string cosmology models and studying inflationary scenarios.  Cosmological observations involving the early ages of universe allow a possibility for testing string theory at energy scales which provide a natural laboratory for stringy physics, see \cite{Tye:2006uv, Cline:2006hu, Kallosh:2007ig, Burgess:2007pz, McAllister:2007bg} for recent reviews on string cosmology.

The renormalization group flow of the gauge theory involves a cascade of Seiberg duality \cite{Seiberg:1994pq} transitions in which the rank of gauge theory drops as $SU(N+M)\times SU(N)\to SU(N-M)\times SU(N)\to SU(N-2M)\times SU(N-M)\to \cdots$.
However, the duality construction assumes that the anomalous mass dimension of the chiral superfields which dictates the renormalization group flow of the couplings in the gauge theory is constant and the same as that in $SU(N)\times SU(N)$, even as the theory flows deep in the infrared after a series of Seiberg duality transitions and the rank of the gauge theory is reduced to say $SU(3M)\times SU(2M)$,  and that the dilaton and the axion are both constants on the gravity side.
Corrections to the anomalous mass dimensions were proposed in  \cite{Hailu:2006uj} where it was shown that including the corrections leads to a consistent picture and also gives novel features on the gravity side which might be useful for constructing cosmological models of the early universe and studying the possibilities for probing distinct stringy signatures.

Here, we present explicit gauge/gravity duality construction which takes the corrections to the anomalous mass dimension on the gauge theory side proposed in \cite{Hailu:2006uj} into account and both the dilaton and the axion run on the gravity side. We find that the sources for the runnings of dilaton and the axion are magnetic couplings of the axion to D7-branes filling 4-d spacetime and wrapping 4-cycles in the throat and invisible Dirac 8-branes which emanate from the D7-branes, as in a Dirac string emanating from a magnetic monopole \cite{Dirac:1931kp,Dirac:1948um}. Duality transitions occur as the theory flows across the wrapped D7-branes and, therefore, the D7-branes also serve as gravitational source of Seiberg duality transformations. The explicit supergravity solutions are found using equations for type IIB flows with $\mathcal{N}=1$ supersymmetry recently obtained in \cite{Hailu:2007ae}. The equations in \cite{Hailu:2007ae} were obtained using $SU(3)$ structures \cite{Grana:2004bg, Grana:2005jc, Dall'Agata:2004dk, Gauntlett:2002sc, Gurrieri:2002wz, LopesCardoso:2002hd, Gauntlett:2003cy}.
Although we emphasize that the corrections to the anomalous dimensions proposed in \cite{Hailu:2006uj} provide a consistent picture and should be good approximations, the solutions we present are general and accommodate other corrections. Our construction provides the first explicit example of a gauge/gravity duality mapping with a running dilaton or a running axion which is an important component towards finding gravity duals to gauge theories with physically more interesting renormalization group flows such as pure confining gauge theories.

The organization of this article is as follows. First we write the corrections to the anomalous mass dimension of the chiral superfields proposed in \cite{Hailu:2006uj} and make a brief discussion. We then move to the supergravity side. We start with writing the equation for the running of the dilaton and reading off from \cite{Hailu:2006uj} a suitable radial variable we will use in analyzing the supergravity system.
We then write a modified and generalized version of the metric and flux ansatz used in the Klebanov-Strassler construction \cite{Klebanov:2000hb} given in \cite{Butti:2004pk, Papadopoulos:2000gj} with additions and changes we make to take the runnings of the dilaton and the axion into account and the corresponding torsion and flux components in $SU(3)$ representations are given in appendix \ref{app-dec}. We then write the set of equations given in \cite{Hailu:2007ae} for a class of flows with imaginary self-dual 3-form flux with running axion and running dilaton, which as we will see is an appropriate class for the supergravity flow we have here. The solutions to all the equations for all the variables in the metric and the flux ansatz are then explicitly written down. Some of the variables in the metric and in the fluxes have the same forms of expressions as in the Klebanov-Strassler solution (except for a redefinition of the radial coordinate) and the solutions for these components are read off from parts of the solutions in \cite{Butti:2004pk}.

We then continue with analyzing the solutions for physical interpretations. With the supergravity solutions in hand, we go backwards and calculate the magnitudes of the anomalous mass dimension predicted by the gravity theory and check that the constant pieces are the same as the values proposed in \cite{Hailu:2006uj}. The gravity side predicts additional nonlinear corrections which depend on the scale of the theory as we expect.
We then show that couplings of the axion to D7-branes filling up 4-d spacetime and wrapping 4-cycles in the extra space and invisible Dirac 8-branes which emanate from the D7-branes are the sources for the runnings of the dilaton and the axion, where we borrow from general discussions on couplings between axion and 7-branes given in \cite{Bergshoeff:2007aa}. One peculiar feature of our solutions is that the $F_1$ flux is not closed. We will explicitly show that this is due to the invisible Dirac 8-branes and a consistent picture emerges starting with the type IIB supergravity action including magnetic couplings of the axion to the D7-branes.
We end with conclusions and discussions.

\section{Renormalization group flow of the gauge theory\label{sec-RGF}}

In this section, we write and discuss the corrections to the anomalous mass dimension of the chiral superfields proposed in \cite{Hailu:2006uj}.
Let us consider $\mathcal{N}=1$ supersymmetric $SU(N+M)\times
SU(N)$ gauge theory with chiral superfields transforming as
$A_1,\,A_2\sim (\Box,\,\bar{\Box})$ and $B_1,\,B_2\sim
(\bar{\Box},\,\Box) $ and the tree level superpotential
\be
W_{\mathrm{tree}}=w\Bigl((A_1 B_1)(A_2 B_2) - (A_1 B_2)(A_2 B_1)
\Bigr), \label{wtreequartic}
\ee
where color indices from the same gauge group are contracted and $w$ is the tree level coupling.
Let us define the classical dimensionless coupling related to
the tree level coupling $w$ by $\eta=\ln (w/\mu^{1+2\gamma_\eta})$, where $\mu$ has a
dimension of mass.
Let the
quantity $\gamma=\gamma_A+\gamma_B$ stand for the anomalous mass
dimension of any one of the meson objects $(A_{i}B_{j})$ made out of the
bifundamental chiral superfields, which contains one $A$ and one $B$
superfields and which must have the same anomalous dimension because
of global symmetries in the theory. Let us denote the
gauge coupling of the larger group, which is $SU(N+M)$ as the theory starts to flow form the edge of the throat, by $g_1$, and that of the smaller group, which is
$SU(N)$ in the same region, by $g_2$, and define \be T_1\equiv 8\pi ^2/g_1^2,\quad T_2\equiv 8\pi ^2/g_2^2.\ee
Suppose we take one of the gauge group factors as weakly coupled
relative to the other, then we can treat one gauge group factor as
a flavor symmetry to the other. The runnings of the physical couplings
 \cite{Shifman:1986zi} with appropriate normalization of the gauge
chiral superfields can then be written as
\be
\beta_{1}=\mu \frac{d T_1(1)}{d\mu}=3(N+M)-2N(1-\gamma(1)),
\label{beta1}
\ee
\be
\beta_{2}=\mu \frac{d T_2(1)}{d\mu}=3N-2(N+M)(1-\gamma(1)),
\label{beta2}
\ee
and
\be
\beta_{\eta}=\mu \frac{d \eta(1)}{d\mu}=1+2\gamma(1).
\label{betaeta}
\ee
The number ``$1$'' in the
parentheses denotes the $l=1$ region, in the ultraviolet region
before the first duality transformation in the cascade.

For the special case of $M=0$, the $SU(N+M)\times SU(N)$ gauge theory reduces to $SU(N)\times SU(N)$ which is dual to type IIB string theory with $N$ number of D3-branes on $AdS_5\times T^{1,1}$ background \cite{Klebanov:1998hh}. In this case, all the three $\beta$ functions given by (\ref{beta1})-(\ref{betaeta}) vanish for the same value of anomalous mass dimension $\gamma=-1/2$ and the theory is conformal. But for $M\ne 0$ there is no common value of anomalous mass dimension for which all the $\beta$ functions vanish and the theory becomes nonconformal.
According to Seiberg duality, $\mathcal{N}=1$ supersymmetric $SU(N)$
electric gauge theory with $N_f \in (3N/2,3N)$ flavors, with appropriate tree level superpotential,  becomes strongly coupled as it flows to the infrared and has a nontrivial conformal
fixed point where it joins a dual infrared free $SU(N_f-N)$ magnetic gauge
theory with $N_f$ flavors.
It was argued in \cite{Klebanov:2000hb} that the $SU(N+M)\times SU(N)$ theory with the quartic tree level superpotential given by (\ref{wtreequartic}) undergoes a series of Seiberg duality transformations,
$SU(N+M)\times SU(N)\to SU(N-M)\times SU(N)\to SU(N-2M)\times SU(N-M)\to \cdots$. However, the same value of $\gamma=1/2$ as in $SU(N)\times SU(N)$ was taken throughout the cascading flow of $SU(N+M)\times SU(N)$.
Analyzing the symmetries and the renormalization group flow in the gauge theory,
it was proposed in \cite{Hailu:2006uj} that
\be
\gamma
(1)=-\frac{1}{2}-\frac{3}{4}\frac{M^2}{N(N+M)}\label{cascs1a}
\ee
and the anomalous dimension for the $l^{th}$ region approaching
the $l^{th}$ step in the duality cascade
involving
$SU(N-(l-2)M)\times SU(N-(l-1)M)$,
\be
\gamma(l)=-\frac{1}{2}-\frac{1}{3}C_{l-2}C_{l-1}= -\frac{1}{2} -
\frac{3M^{2}}{4(N+2M-lM)(N+M-lM)},\label{cascsl}
\ee
where
\be
C_{l}\equiv \frac{3}{2}\frac{M}{N-l M}. \label{cldef}
\ee

These values of anomalous mass dimensions pass various consistency checks. Let us review some of the main reasons why these corrections are physically sensible.\footnote{The exact value of anomalous mass dimension would be nonlinear and a function of the scale of the theory itself. What we are emphasizing here is that first, corrections are necessary and need to be included and second, the corrections proposed in \cite{Hailu:2006uj} are reasonable and probably even the best possible constant values of the anomalous mass dimension. Nevertheless, the supergravity solutions we present here are general and accommodate other corrections. With the supergravity solutions in hand in section \ref{sec-solns}, we will go backwards and find gravity predictions for the nonlinear corrections to the anomalous dimension in section \ref{sec-amdgs}.}
\begin{itemize}
\item First we like to repeat stating that corrections are necessary simply because $SU(N+M)\times SU(N)\ne SU(N)\times SU(N)$.
\item Symmetries: The $M^{2}/(N+2M-lM)(N+M-lM)$ in the corrections given in (\ref{cascsl}) obeys the symmetries ($M\to -M$ and $N\to N-(2l-3)M$) in the $SU(N-(l-2)M)\times SU(N-(l-1)M)$ theory and it is a unique combination.
\item The magnitudes of the corrections are within range for Seiberg duality transformations. The mass dimension $d[\mathcal{O}]$ of the operator in the tree level superpotential given by (\ref{wtreequartic}) $\mathcal{O}\equiv(A_1 B_1)(A_2 B_2) - (A_1 B_2)(A_2 B_1)$ for the flow between the $l^{th}$ and the $(l+1)^{th}$ duality transitions is
\be
d[\mathcal{O}]=4+2\gamma=
3 -
\frac{3M^{2}}{2(N-lM)(N+M-lM)}.\label{dop-1}
\ee
For the flow involving $\frac{N}{M}-1$ duality transitions, we have
$\frac{9}{4}\le d[\mathcal{O}] \le 3-\frac{3}{2}\frac{M^2}{N(N+M)}$. For the mesons, $\frac{9}{8}\le d[(A_iB_j)] \le \frac{3}{2}-\frac{3}{4}\frac{M^2}{N(N+M)}$. Thus the operator $\mathcal{O}$ is relevant throughout the flow. Note that the mesons have mass dimension $d[(A_iB_j)] \ge \frac{9}{8}$ and this is consistent with the $\ge 1$ bound for the mass dimension of scalars at a conformal fixed point. Here the theory flows nearby such a fixed point. The mass dimension  $d[(A_iB_j)]$ changes from $\frac{11}{8}$ to $\frac{9}{8}$ across the last $(\frac{N}{M}-1)^{th}$ duality transition and a further duality transformation would have made the mass dimension of the mesons $<1$ which would be inconsistent. Here, $d[(A_iB_j)]$ actually diverges for $l=N/M$ and the duality cascade has to end with $SU(2M)\times SU(M)$ and has only $K-1$ transitions for $N$ an integral $K$ multiple of $M$.
\item That the duality transitions end in $SU(2M)\times SU(M)$ for $N$ an integral multiple of $M$ is what we expect from Seiberg's picture of electric-magnetic duality \cite{Seiberg:1994pq} in which a strongly coupled $\mathcal{N}=1$ supersymmetric $SU(N)$ electric gauge theory with $N_f$ flavors with  appropriate tree level superpotential deformation has dual magnetic description in the so called conformal window where the number of flavors is within $3N/2< N_f<3N$ and in the free magnetic phase where $N+2\le N_f \le 3N/2$. Thus $SU(2M)\times SU(M)$, which contains a similar renormalization group flow as that for $SU(2M)$ with $2M$ flavors falls outside the duality window and cannot be expected to undergo a duality transition. Indeed, $SU(2M)$ with $2M$ flavors has a quantum deformed moduli space of vacua \cite{Seiberg:1994bz}.
\end{itemize}

Nevertheless, we would like to be more general and treat the $C_l$ in (\ref{cascsl}) as parameters which could take other values than (\ref{cldef}) and the supergravity solutions are written in terms of these general parameters. What is physically necessary is that the magnitudes are right for making the operator in the tree level superpotential relevant.

\section{Supergravity background}

The IIB background geometry is $AdS_5\times T^{1,1}$, where $T^{1,1}=S^3\times S^2$ is the coset space of ${SU(2)\times SU(2)}/{U(1)}$. Let us denote the 2- and the 3-cycles associated with $S^2$ and $S^3$ respectively by $\omega_2$ and $\omega_3$. The radius of the $AdS_5$ factor is parameterized by $r$ and the background geometry is a cone with $T^{1,1}$ base. The $M$ number of D5-branes, on the gauge theory side, fill up 4-d spacetime and wrap $S^2$ near the tip of the cone and become fractional D3-branes which give rise to 3-form R-R $F_3$ flux on the gravity side after geometric transition. The $N$ number of regular D3-branes give a self-dual 5-form R-R flux $(1+\star)F_5$. The interplay between these fluxes creates a backreaction NS-NS 2-form potential with corresponding 3-form flux $H_3$. Let us denote the radial location where the flow corresponding to the gauge theory with rank $SU(N+M)\times SU(N)$ starts from the edge of the throat by $r=r_0$ and the location where the $l^{th}$ duality transition takes place by $r_l$. The throat is glued to a larger background beyond $r=r_0$ and the value of $r$ decreases as the theory flows down the throat. The explicit supergravity solutions in the throat with the corrections to the anomalous mass dimension included on the gauge theory side will be written down and we will see that D7-branes filling 4-d spacetime and wrapping 4-cycle at each $r_l$ source the runnings of the dilaton and the axion.

\subsection{Running dilaton\label{sec-dm}}

First, in this section, we write the equation for the running of the dilaton obtained in \cite{Hailu:2006uj} due to the corrections to the anomalous mass dimension. We will also look at the strength in the running of the backreaction NS-NS potential and read off appropriate radial variable  we will use in finding the supergravity solutions in section \ref{sec-solns}.

The effective gauge coupling in the gauge theory is parameterized by
\be
T_+\equiv T_1+T_2
\ee
and this is the variable which needs to get mapped to the string coupling. The running of $T_+$ translates into a running of the string coupling through a running dilaton $\Phi$. The mapping between the couplings in the gauge and in the gravity theories is
\be
T_+ =  \frac{2\pi}{g_s e^\Phi}, \label{t1p2phi}
\ee
where $g_s$ is the bare string coupling.
The difference in the couplings in the gauge theory
\be
T_-\equiv T_1-T_2
\ee
drives a backreaction NS-NS potential in the gravity theory and dictates the strength of its running  \cite{Klebanov:1999rd,Morrison:1998cs}. A guide in how $T_{-}$ translates to on the gravity side is to recall that the NS-NS 2-form potential $B_2\sim h_1 \omega_2$ for the case of the singular conifold solution in \cite{Klebanov:2000hb,Klebanov:2000nc,Klebanov:1999rd, Gubser:1998fp,  Klebanov:1998hh} is \footnote{For now, we will use this mapping only as guide to write appropriate radial variables for the flows between duality transitions in the supergravity theory.  We will calculate the anomalous mass dimensions the supergravity solutions predict such that this mapping holds in section \ref{sec-amdgs}.  The variable $h_1$ is in anticipation of the notation in the NS-NS 3-form flux in section \ref{sec-ansatz}.}
\be
 T_-=  \frac{2\pi}{g_s e^\Phi}({\hat{ b}-1})
 =\frac{2\pi}{g_s e^\Phi}({\bar {b}_2(\mathrm{mod}\,2))}\label{t1m2phi}
  \ee
where
 \be
  \qquad  \hat b-1 = \bar{b}_{2}  \, (\mathrm{mod}\, 2),\qquad \bar{b}_2 =  b_2 -1,
  \ee
and
\be
 b_2\equiv - \frac{1}{\pi^2 \alpha'}h_1\int_{S^2}\omega_2. \label{b2def}
\ee
The variable $\bar{b}_2$ measures the deviation of $T_{-}$ from zero due to the backreaction potential from where the two gauge couplings have equal magnitude in $r_{1} < r < r_{0}$. The supergravity flow can be arranged such that $b_2(r)$ vanishes at the edge of the throat, $r=r_0$, and $0 \le  \hat b  \le 2$. The scale in the gauge theory is mapped to the $AdS_5$ radius as \be \frac{\Lambda}{\Lambda_c}=\frac{r}{r_0},\ee where $\Lambda_c$ is the ultraviolet scale of the gauge theory at the edge of the throat where $r=r_0$.

It was obtained in \cite{Hailu:2006uj} that
\be
\frac{d}{d\ln {(r/r_0)}} \,e^{-\Phi}=-S_{l},\label{t1p2phie}
\ee
\be
\frac{d}{d\ln {(r/r_0)}} \Bigl(e^{-\Phi} {\bar {b}_2 } \Bigr) = D_l,
\label{t1m2phie}
\ee
where two sets of
constant parameters were introduced,
\be
S_{l}\equiv \bar{s}_l \frac{g_s M}{ \pi}\equiv s_l \,g_s,\quad {} D_l\equiv
(3+d_l)\frac{g_s M}{\pi},\label{DlSl}\ee with
\be \bar{s}_l\equiv \frac{1}{2}(C_{l-1}+C_{l-2}),\quad d_l\equiv \frac{1}{2}(C_{l-1}-C_{l-2}).\label{sdcl-def}
\ee
The Klebanov-Strassler construction corresponds to $s_l=d_l=0$ and, in this case, the dilaton is constant.

Because the magnitudes of the coefficients $s_l$ and $d_l$ change after each duality transition, the dilaton and the backreaction potential run with different strengths across locations of duality transformations. We find it convenient to define
\be
\tau_l\equiv (3+d_l)t,\quad t \equiv \ln(r/r_{c}),\label{taut-not}
\ee
where $r_c$ is the $AdS_5$ radius at which the $F_5$ flux would vanish.
The equations (\ref{t1p2phie}) and (\ref{t1m2phie}) then become\footnote{In order not to clutter the text, we drop off the index $l$ in all symbols with the understanding that the equations and expressions we write are for the flow between the $(l-1)^{th}$ and the $l^{th}$ locations of duality transitions. We put the index only when we need it in our discussions.}
\be
\frac{d}{d\tau} \,e^{-\Phi}=-g_s \frac{s}{3+d},\label{t1p2phietau}
\ee
\be
\frac{d}{d\tau} \Bigl(e^{-\Phi} {\bar {b}_2 } \Bigr) = \frac{g_s M}{ \pi}.
\label{t1m2phietau}
\ee

The main result we take here from \cite{Hailu:2006uj} is the equation for the running of the dilaton given by (\ref{t1p2phietau}). We also see the appropriate radial variable is $\tau$, since the strength of the running of the backreaction potential and, consequently the fluxes, is different between different locations of duality transitions. The complete set of supergravity solutions consistent with the running of the dilaton will be obtained in section \ref{sec-solns}.

\subsection{Metric and flux ansatz\label{sec-ansatz}}

We will see later that the supergravity background we have here is accommodated in the class of supergravity flows with imaginary self-dual 3-form flux given in \cite{Hailu:2006uj}.
The metric and flux ansatz we use is a generalized version of that in the Klebanov-Strassler solution as written in \cite{Butti:2004pk, Papadopoulos:2000gj} with additions and modifications we make to accommodate the runnings of the dilaton and the axion we have here. We work with the supergravity action in the string frame.\footnote{All our calculations in this article are in the string frame. The metric in the string frame $G_{MN}(string)$ and the metric in the Einstein frame $G_{MN}(Einstein)$ are related as $G_{MN}(string)=e^{\Phi/2}G_{MN}(Einstein)$.} Now we write the metric and flux ansatz in \cite{Butti:2004pk, Papadopoulos:2000gj}.\footnote{The $\chi$ in the $H_3$ flux in \cite{Papadopoulos:2000gj} and \cite{Butti:2004pk} which was useful for interpolating flow between the Klebanov-Strassler and the Maldacena-Nunez solutions is not needed here and is set to zero.} For the metric,
\begin{equation}
ds^2=e^{2A(\tau)}\eta_{\mu \nu} dx^\mu dx^\nu + ds_6^2(y),\label{10dmetric2}
\end{equation}
where
\be\label{ds6s-1}
ds_6^2=\delta_{mn}G^m G^n=\delta_{i\bar{j}}Z^i\bar{Z}^{\bar{j}},
\ee
with
\be\label{ZiGi}
Z^1=G^1+i G^2,\quad Z^2=G^3+i G^4,\quad Z^3=G^5+i G^6.
\ee
and their complex conjugates $\bar{Z}^{\bar{i}}=({Z}^{i})^\ast$.
The $G^m$ are real differential 1-forms which are expressed in terms of linear combinations of the coordinate 1-forms $dy^n$ on $Y$ with coefficients which are functions of $y$,
\ba
G^1=E^1,\quad G^2=\mathcal{A} E^2+\mathcal{B} E^4,\quad G^3=E^3,\nonumber \\G^4=\mathcal{B} E^2-\mathcal{A} E^4,\quad G^5=E^5,\quad G^6=E^6,
\ea
where
\be
\mathcal{A}^2+\mathcal{B}^2=1,
\ee
\begin{eqnarray}&
E^1=e^{\frac{x+g}{2}}e_1,\quad E^2=e^{\frac{x+g}{2}}e_2,\quad
E^3=e^{\frac{x-g}{2}}\tilde{\epsilon_1},\quad
E^4=e^{\frac{x-g}{2}}\tilde{\epsilon_2},&\nonumber\\
&E^5=e^{\frac{-6p-x}{2}}d\tau,\quad
E^6=e^{\frac{-6p-x}{2}}\tilde{\epsilon_3}&\label{Ei}
\end{eqnarray}
and
\begin{eqnarray}&
e_1=d\theta_1,\quad e_2=-\sin{\theta_1}d\phi_1,\quad
\tilde{\epsilon_1}=\epsilon_1-a e_1,\quad
\tilde{\epsilon_2}=\epsilon_2-a e_2, &\nonumber\\&
\tilde{\epsilon_3}=\epsilon_3+\cos \theta_1 d\phi_1,\quad
\epsilon_1=\sin \psi \sin\theta_2 d\phi_2+\cos\psi d\theta_2,
&\nonumber\\& \epsilon_2=\cos \psi \sin\theta_2 d\phi_2-\sin\psi
d\theta_2,\quad \epsilon_3=d\psi+\cos\theta_2
d\phi_2.&\label{epsiloni}
\end{eqnarray}
where $\theta_1,\,\theta_2\, \in [0,\pi]$, $\phi_1,\,\phi_2\, \in [0,2\pi]$, and $\psi\, \in [0,4\pi]$.
The ansatz for the fluxes is\footnote{The ansatz for the $F_1$ flux is based on calculation we do in section \ref{sec-solns}. We have also added a factor of $e^{-\Phi}$ in the ansatz for $F_3$ so that we will have $dF_3=-F_1\wedge H_3$ after using relations between $d\Phi$ and $F_1$ and between $F_3$ and $H_3$ as we will see in section \ref{sec-eqns}.}
\begin{eqnarray}
H_3&=&d\tau\wedge \Bigl(h_1'(\epsilon_1\wedge \epsilon_2+e_1\wedge
e_2) +h_2'(\epsilon_1\wedge e_2-\epsilon_2\wedge
e_1)\Bigl)\nonumber\\
&&+h_2\tilde{\epsilon}_3\wedge (\epsilon_1\wedge
e_1+\epsilon_2\wedge e_2),\label{fluxes}
\end{eqnarray}
\begin{eqnarray}
F_3&=& e^{-\Phi} P  \Bigl(\tilde{\epsilon}_3\wedge (\epsilon_1\wedge
\epsilon_2+e_1\wedge e_2-b (\epsilon_1\wedge e_2-\epsilon_2\wedge
e_1))\nonumber\\&&+b' d\tau\wedge (\epsilon_1\wedge
e_1+\epsilon_2\wedge e_2)\Bigr),\label{F3-ansatz}
\end{eqnarray}
\begin{eqnarray}
F_5=K\,e_1\wedge e_2 \wedge \epsilon_1 \wedge \epsilon_2
\wedge \epsilon_3,
\quad
K=2e^{-\Phi}P(h_1+bh_2), \label{khh1h2}
\end{eqnarray}
\be
F_{1}=\frac{s}{3+d}\,\tilde{\epsilon}_3.\label{F1-ansatz}
\ee

Sometimes we use the notation used in the Klebanov-Strassler solution for the warp factor, $h$, which is related to $A$ as $h=e^{-4A}$.
The parameters in the metric and in the fluxes above, $A, \mathcal{A}$, $\mathcal{B}$, $a$ $x$, $g$, $p$, $h_1$, $h_2$, $b$, and $K$ are functions of $\tau$, where $\tau$ is now the variable defined in (\ref{taut-not}) with the magnitudes of $s$ and $d$ in the coefficients of $t$ changing after every duality transition. The symbol $P$ is constant and proportional to the number of regular D3-branes with $P\equiv-\alpha' M/4$ and $K$ describes the effective number of D3-branes as a function of $\tau$ which decreases as the theory flows down towards the bottom of the throat.  The solutions for the variables in the 3-form fluxes will be related such that imaginary self-dual combination is formed. The form of the $F_1$ flux is taken based on the running of the dilaton and the imaginary self-dual combination of the 3-form fluxes. Note that the $F_1$ flux is not closed because $\tilde{\epsilon}_3=d\psi+\cos\theta_1 d\phi_1+\cos\theta_2 d\phi_2$. In addition, $F_1$ changes across locations of duality transitions, since the magnitudes of $s$ and $d$ change. We will see that these are due to D7-branes and couplings of the axion to D7-branes via invisible Dirac 8-branes.

\subsection{Torsion and flux components\label{sec-ftc}}

Here we summarize the decomposition of the torsion and the fluxes in $SU(3)$ representation. See \cite{Hailu:2007ae} for details of our notations.
We write the fundamental 2-form $J$ and holomorphic 3-form $\Omega$ as
\begin{eqnarray}\label{JZ-1}
J&=& \frac{1}{2}J_{i\bar{j}}\,{Z^{i}}{ {\wedge} }{\bar{Z}^{\bar{j}}}=
\frac{i}{2} \delta_{i\bar{j}}{Z^{i}}{ {\wedge} }{\bar{Z}^{\bar{j}}},
\end{eqnarray}
\begin{eqnarray}\label{OmZ-1}
\Omega &=&\frac{1}{6}\Omega_{ijk}{Z^i}{ {\wedge} }{Z^j}{ {\wedge} }{Z^k}=\frac{1}{6}\epsilon_{ijk}{Z^i}{ {\wedge} }{Z^j}{ {\wedge} }{Z^k}=
{Z^1}{ {\wedge} }{Z^2}{ {\wedge} }{Z^3}.
\end{eqnarray}
The torsion components come in the variations of $J$ and $\Omega$,
\be\label{dJ-1a}
dJ=-\frac{3}{2}\mathrm{Im}(W_1^{(1)} \bar{\Omega})+(W_4^{(3)}+W_4^{(\bar{3})})\wedge J+(W_3^{(6)}+W_3^{(\bar{6})}),
\ee
\be\label{dOm-a1a}
d\Omega=W_1^{(1)} J^2+W_2^{(8)}\wedge J+W_5^{\bar{3}}\wedge \Omega.
\ee
The 3-form fluxes, $F_3$ and $H_3$, have only internal components with $(0,3)$ and $(1,2)$ forms and their conjugates and are decomposed as $dJ$,
\be\label{H3-1a}
H_3=-\frac{3}{2}\mathrm{Im}(H_3^{(1)} \bar{\Omega})+(H_3^{(3)}+H_3^{(\bar{3})})\wedge J+(H_3^{(6)}+H_3^{(\bar{6})}),
\ee
\be\label{F3-1a}
F_3=-\frac{3}{2}\mathrm{Im}(F_3^{(1)} \bar{\Omega})+(F_3^{(3)}+F_3^{(\bar{3})})\wedge J+(F_3^{(6)}+H_3^{(\bar{6})}).
\ee
The $F_5$ part of the self-dual 5-form flux $(1+\star)F_5$ has only internal components and is written as
\be
{F_5}=(F_5^{(3)}+F_5^{(\bar{3})})\wedge J\wedge J.
\ee
The explicit expressions of the torsion and flux components in $SU(3)$ representations for the metric and flux ansatz in section \ref{sec-ansatz} are given in appendix \ref{app-dec}.

\subsection{Equations\label{sec-eqns}}

Now we write the equations we need to solve in finding the supergravity solutions with the corrections to the anomalous mass dimension included in the gauge theory and the metric and flux ansatz in section \ref{sec-ansatz}. The first important feature of the background we have here is the fact that the corrections to the anomalous mass dimension make the dilaton run.  Now if the axion is constant, we know from the discussions in \cite{Hailu:2007ae} that the 3-form fluxes need to have nonprimitive components. On the other hand, it was shown in \cite{Hailu:2007ae} that there is a class of flows with imaginary self-dual 3-form flux and a running dilaton if the axion also runs at the same time such that $-i\bar{\partial}\Phi e^{-\Phi}/g_s+ F_1^{(\bar{3})}=0$.\footnote{This implies that the dilaton-axion coupling coefficient $\frac{i}{g_s} e^{-\Phi}+C_0$ is constant if the $F_1$ flux is closed as discussed in \cite{Hailu:2007ae}.} Indeed, we find that this class of flows has suitable features to describe the supergravity side.
The running of the dilaton is balanced by the $F_1$ flux and these also lead to promotion of the  components of the fluxes and metric in the Klebanov-Strassler solution to more general ones which are balanced by a corresponding promotion of the torsion components in the $3\oplus \bar{3}$ sector.

Let us write the equations which relate the fluxes, the torsion and the runnings of the dilaton and the axion given for the specific class of flows with imaginary self-dual 3-form flux
in \cite{Hailu:2007ae}, where more details can be found.
For the components in the $3\oplus \bar{3}$ sector, the equations are
\be
F_3^{(\bar{3})}=H_3^{(\bar{3})}=0,
\quad
g_s e^{\Phi}F_5^{(\bar{3})}=-4i\bar{\partial}\ln \alpha+\frac{1}{2}g_s e^{\Phi}F_1^{(\bar{3})},\label{F5F1}
\ee
\be
\quad
W_4^{(\bar{3})}=-4\bar{\partial}\ln \alpha-ig_s  e^{\Phi}F_1^{(\bar{3})},\quad W_5^{(\bar{3})}=-6\bar{\partial}\ln \alpha,\label{W4W5F1}
\ee
\be
\bar{\partial}\Phi=-i g_s e^{\Phi}F_1^{(\bar{3})},\quad \bar{\partial}A=2\bar{\partial}\ln \alpha.\label{dAF1isd}
\ee
In addition, the equations in the $6\oplus\bar{6}$ sector are
\be H_3^{(6)}= g_s  e^{\Phi} \star_6 F_3^{(6)},\qquad  W_3^{(6)}=0\label{H3F3W3}\ee
and for the $1\oplus1$ and the $8\oplus8$ sectors,
\be
W^{(1)}_1=0,\quad W^{(8)}_2=0,\quad H^{(1)}_3=0,\quad F^{(1)}_3=0.\label{W1W2H31F31}
\ee

Now let us see an implication of the factor of $e^{-\Phi}$ we put in the ansatz for $F_3$. We see that $dF_3=d(e^{-\Phi}/g_s)\wedge g_s e^\Phi F_3=F_1\wedge g_s e^\Phi i F_3=-F_1\wedge g_s e^\Phi \star_6 F_3=-F_1\wedge H_3$ which is what we want with the running axion we have here.

\subsection{Solutions\label{sec-solns}}

Now we want to find the supergravity solutions for all the variables in the metric and in the fluxes.\footnote{The corrections to the anomalous dimension are encoded in and the solutions are expressed in terms of $s$, $d$, and $\tau=(3+d)t$. The values of $s$ and $d$ are given by (\ref{sdcl-def}), (\ref{DlSl}) and (\ref{cldef}) for the proposal in \cite{Hailu:2006uj}. The supergravity solutions we present here are general and accommodate other possible values of $s$ and $d$ too. All components of the solutions contain (and reduce to) the ones in the Klebanov-Strassler solution for $s=d=0$.}
First, the equation for the running of the dilaton given by (\ref{t1p2phietau}) gives the solution\footnote{The solutions for many of the variables are written in such integral form with the understanding that starting with the value of a variable at some radial location in the flow the value at another location is obtained by doing the integration over the corresponding radial range with the different corresponding values of $s$, $d$ and $\tau$ taken for the flows between different locations of duality transitions.}
\be
e^{-\Phi}=-\int^\tau g_s\frac{s}{3+d}d\tau.\label{phi-soln}
\ee
The components of the torsion and the fluxes in  $SU(3)$ representations are given in appendix \ref{app-dec}.
The equations in the $1\oplus1$, the $8\oplus8$ and the $6\oplus\bar{6}$ sectors have the same form  as that studied in \cite{Butti:2004pk} with $\chi$ set to zero in the ansatz for $H_3$ and the relations and solutions which come from these sectors are read off from \cite{Butti:2004pk}. The equations $W_1^{(1)}=0$ and $W_2^{(8)}=0$ in (\ref{W1W2H31F31}) together with the expressions given by (\ref{W11bZ}) and (\ref{W2Z}) give the relations
\be
e^{2g}=-1-a^2-2a \cosh \tau,\label{gsol0}\ee
\be
\mathcal{A}=(\cosh \tau +a) \csch \tau,\quad \mathcal{B}=e^{g} \csch \tau,\label{Us0}
\ee
The vanishing of $F_3^{(1)}$ and $H_3^{(1)}$  in (\ref{W1W2H31F31}) together with the expressions given by (\ref{F31}) and (\ref{H31Z}) give the relations
\be b=-\tau \csch \tau,\label{b-sol}
\ee
\be
h_{1}=h_{2} \cosh \tau.\label{h1-soln1}
\ee
The equations $F_3^{(\bar{3})}=0$ and $H_3^{(\bar{3})}=0$ in (\ref{F5F1}) with (\ref{F33}), (\ref{H33bZ}) and the above relations, then, give
\be
a=-\sech \tau.\label{asol0}
\ee
With (\ref{asol0}), the relations in (\ref{gsol0}) and (\ref{Us0}) become
\be
e^{g}=\tanh \tau,\label{gsol0b}
\ee
\be
\mathcal{A}=\tanh \tau,\quad \mathcal{B}=\sech \tau,\label{Us0b}
\ee
The equations in the $6\oplus\bar{6}$ sector given in (\ref{H3F3W3}) with the expressions given by (\ref{W36bZ}), (\ref{H36bZ}) and (\ref{F36bZ}) and the above relations, then, give
\be
h_2=\frac{1}{4} g_sM\alpha'(1-\tau \coth \tau)\csch \tau.\label{h2-soln}
\ee
Note that the $e^{-\Phi}$ factor in the ansatz for $F_3$ which was useful in obtaining $dF_3=-F_1\wedge H_3$ is useful here too in that $h_2$ now does not explicitly contain the dilaton factor which appears in the solutions in \cite{Butti:2004pk} and the equation (\ref{H3F3W3}) in the $6\oplus \bar{6}$ sector is satisfied.
Moreover, the Bianchi identity for the 5-form flux given in (\ref{khh1h2}) with (\ref{h1-soln1}) and (\ref{h2-soln}) gives
\be
K=\frac{1}{2}e^{-\Phi}M\alpha'(\tau\csch \tau-\cosh \tau) h_2\label{Ksolna}
\ee
The expressions (\ref{b-sol})-(\ref{Ksolna}) are the same as in the Klebanov-Strassler solution  written in the form given in \cite{Butti:2004pk} with nonzero $\Phi$ and where $\tau$ now is as defined in (\ref{taut-not}).

Next let us consider the remaining equations in the $3\oplus \bar{3}$ sector.  Using the expression for the running of the dilaton given by (\ref{t1p2phietau}), the equation for the running of the dilaton given in (\ref{dAF1isd}) and the metric, we obtain the $\bar{3}$ component of the $F_1$ flux,
\be
F_{1}^{(\bar{3})}=\frac{i}{2}\frac{s}{3+d} e^{\frac{6p+x}{2}}\bar{Z}^{\bar{3}}=\frac{i}{2}\frac{s}{3+d} e^{\frac{6p+x}{2}}\bar{Z}^{\bar{3}}.
\ee
The total $F_1$ flux is then\footnote{This expression is the reason we wrote the ansatz for the $F_1$ flux as given by (\ref{F1-ansatz}).}
\be
F_1=F_{1}^{(3)}+F_{1}^{(\bar{3})}=\frac{s}{3+d} e^{\frac{6p+x}{2}}G^6=\frac{s}{3+d}\tilde{\epsilon}_3.\label{F1-tot1}
\ee
Then combining the two equations in (\ref{W4W5F1}), $3W_4^{(\bar{3})}-2W_5^{(\bar{3})}=-3 i g_s e^\Phi F_1^{(\bar{3})}$, with (\ref{W43bZ}), (\ref{W53bZ}) and (\ref{asol0})-(\ref{Us0b}) gives
\be
2(\coth \tau+3p'+x')-3e^{-6p-2x}=\frac{3s}{3+d}g_s e^\Phi.\label{v0}
\ee
Moreover, the equation given by (\ref{F5F1}), $g_s e^{\Phi}(F_5^{(\bar{3})}-\frac{1}{2}F_1^{(\bar{3})})=-4i\bar{\partial}\ln \alpha$, with (\ref{asol0})-(\ref{Us0b}) in (\ref{F53b}) for $F_5^{(\bar{3})}$ gives
\be
\bar{\partial}\ln\alpha=(\frac{1}{64}g_s e^{\Phi+3p-\frac{3}{2}x}K +\frac{1}{16}\frac{s}{3+d} g_s e^{\Phi+3p+\frac{1}{2}x})\bar{Z}^{\bar{3}}\label{dalpab0}
\ee
and the equation
$W_5^{(\bar{3})}=-6\bar{\partial}\ln\alpha$ in (\ref{W4W5F1}) with (\ref{W53bZ}) and (\ref{dalpab0}) gives
\be
2\coth \tau +6 p'-x'-\frac{3}{8}g_s e^{\Phi-2x} K =\frac{3}{2}\frac{s}{3+d} g_s e^\Phi.\label{u0}
\ee
Now let us define
\be v\equiv e^{6p+2x}\qquad\mathrm{and}\qquad u\equiv e^{2x}\label{uvdefn}\ee
and then we have from (\ref{v0}),
\be\label{ve-1}
v'+(2\coth \tau-\frac{3s}{3+d} g_s  e^\Phi)v-3=0
\ee
and from (\ref{u0}), (\ref{ve-1}) and noting $-(\frac{4}{3}\coth \tau+\frac{2}{3}\frac{v'}{v}-\frac{s}{3+d}g_s  e^\Phi)=-(\frac{2}{v}+\frac{s}{3+d}g_s  e^\Phi)$,
\be
u' -(\frac{2}{v}+\frac{s}{3+d}g_s  e^\Phi)u+\frac{1}{4}g_s e^{\Phi}K=0.\label{u-eq1}
\ee
For the warp factor, we have from (\ref{dAF1isd}), (\ref{dalpab0}), (\ref{uvdefn}), and the metric (\ref{10dmetric2}),
\be
A'=g_s e^\Phi \left(\frac{1}{16}\frac{K}{u}+\frac{1}{4}\frac{s}{3+d}\right).\label{A-eq1}
\ee
The solutions for $v$, $u$, and $A$ are\footnote{Note that $v$ is fixed first using $f_1$ given by (\ref{f1-def}) in (\ref{v-soln}), the resulting $v$ is used in (\ref{f2-def}) to fix $f_2$, then the resulting $f_2$ and $f_3$ given by (\ref{f3-def}) are used in (\ref{u-soln}) to find $u$, and finally the resulting $u$ is used in (\ref{f4-def}) to obtain $f_4$ which is then used in (\ref{Ah-soln}) to find $A$. Alternatively, it may be more convenient to work with equations (\ref{ve-1}), (\ref{uh-sola}) and (\ref{h-eqa}) as discussed in the paragraph containing  equations (\ref{uh-sola}) and (\ref{h-eqa}).}
\be
v(\tau)=3e^{-\int^\tau f_1(\tau')d\tau'}\int^\tau \left(e^{\int^{\tau''}f_1(\tau''')d\tau'''}\right)d\tau'',\label{v-soln}
\ee
\be
u(\tau)=e^{-\int^\tau f_2(\tau')d\tau'}\int^\tau \left(e^{\int^{\tau''}f_2(\tau''')d\tau'''}f_3(\tau'')\right)d\tau'',\label{u-soln}
\ee
\be
A(\tau)={\int^\tau f_4(\tau')d\tau'}\qquad \mathrm{or} \qquad h(\tau)=e^{-{4}{\int^\tau f_4(\tau')d\tau'}} ,\label{Ah-soln}
\ee
where
\be
f_1\equiv 2\coth \tau-\frac{3s}{3+d}g_s e^\Phi,\label{f1-def}
\ee
\be
f_2\equiv -(\frac{2}{v}+\frac{s}{3+d}g_s  e^\Phi),\label{f2-def}
\ee
\be
f_3\equiv\frac{1}{4}g_s e^{\Phi}K,\label{f3-def}
\ee
\be
f_4\equiv g_s e^\Phi \left(\frac{1}{16}\frac{K}{u}+\frac{1}{4}\frac{s}{3+d}\right)\label{f4-def}
\ee
with $\Phi$ given by (\ref{phi-soln}) and $K$ given by (\ref{Ksolna}). A consistent set of boundary conditions we need are $v(0)=0$ and $u(0)=0$.
Sometimes we will write the $F_1$ flux given by (\ref{F1-tot1}) in terms of $u$ and $v$ as
\be
F_1=\frac{s}{3+d} \frac{v^{1/2}}{u^{1/4}}G^6.\label{F1-tot2}
\ee
Thus we have found the solutions to all the variables in the metric and flux ansatz and all the equations are solved.

Note that equation (\ref{ve-1}) can be easily solved, or (\ref{v-soln}) can be easily integrated to find $v$ and can rewrite the remaining two equations for $u$ and $A$ given by (\ref{u-eq1}) and (\ref{A-eq1}) in slightly more convenient forms.
Dividing (\ref{u-eq1}) by $u$ and using (\ref{A-eq1}) and $h=e^{-4A}$, we can write
\be
\frac{d}{d\tau}\ln(\frac{u}{h})=\frac{2}{v}+\frac{2s}{3+d}g_s e^\Phi.\label{uh-sola}
\ee
We can also rewrite (\ref{A-eq1}),
\be
h'+ \frac{1}{4}g_s e^\Phi K\frac{h}{u}+ g_s e^\Phi\frac{s}{3+d}h=0\label{h-eqa}
\ee
and we use the $h/u$ obtained in (\ref{uh-sola}) in the second term and solve (\ref{h-eqa}) for $h$ which is then used with the expression for $u/h$ obtained from (\ref{uh-sola}) to obtain $u$.

\subsection{Gravity theory prediction of the anomalous mass dimension\label{sec-amdgs}}

Now, with the supergravity solutions in hand, we can find the magnitudes of the corrections to the anomalous mass dimensions that the gravity side predicts assuming that the mapping given by (\ref{t1m2phi}) holds.
First let us see how the solutions in section \ref{sec-solns} and the mapping between the running of the gauge couplings and the running of the dilaton and the periodic NS-NS potential through $\omega_2$ in the gravity theory match. The corresponding component in the $H_3$ flux given in (\ref{fluxes}) is
\be h_1'\, d\tau\wedge (\epsilon_1\wedge \epsilon_2+e_1\wedge e_2)=-2 h_1'\,  d\tau\wedge \omega_2,\ee  where
\be
\omega_2=-\frac{1}{2}(\epsilon_1\wedge \epsilon_2+e_1\wedge e_2)=\frac{1}{2} (\sin \theta_1 d\theta_1\wedge d\phi_1 -\sin \theta_2 d\theta_2 \wedge d\phi_2)
\ee
and $\int_{S^2}\omega_2=4\pi$.
This with (\ref{b2def}), (\ref{h1-soln1}) and (\ref{h2-soln}) gives
\be
b_2=-\frac{4}{\pi \alpha'} h_1=- \frac{g_s M}{\pi} (1-\tau \coth \tau)\coth \tau
\ee
and with the form of ansatz for $H_3$ and $F_3$ in (\ref{fluxes}) and (\ref{F3-ansatz}) where $e^{-\Phi}$ is put in $F_3$, we have
\ba
\frac{d}{dt}  {\bar {b}_2 } &=& (3+d)\frac{d}{d\tau} {\bar {b}_2 } =-(3+d)\frac{g_s M}{\pi} \frac{d}{d\tau} \Bigl((1-\tau \coth \tau)\coth \tau\Bigr)\nn\\&=&\frac{g_s M}{\pi}(3+d)\Bigl(1 + 2(1 -  \tau \coth \tau) {\csch}^2 \tau\Bigr).\label{db2epht}
\ea
The running of the dilaton obeys (\ref{t1p2phietau}) or (\ref{t1p2phie}).
Now we can find the magnitudes of the anomalous mass dimensions implied by the supergravity solutions.
Note that comparing (\ref{t1m2phie}) and (\ref{db2epht}), we see that the gravity side involves promoting the $d_l$, where we put back the $l$ indices, in (\ref{sdcl-def}) to $\bar{d}_l(\tau_l)$ such that
\be
(3+\bar{d}_l(\tau_l)) = (3+d_l)\Bigl(1 + 2(1 -  \tau_l \coth \tau_l) {\csch}^2 \tau_l\Bigr)\label{dltaudl}
\ee
while the $\bar{s}_l$ stay the same constants given by (\ref{sdcl-def}). Thus we have from  (\ref{dltaudl}) that
\be
\bar{d}_l(\tau_l)=d_l+ 2(3+d_l)(1 -  \tau_l \coth \tau_l) {\csch}^2 \tau_l
\ee
which amounts to promoting the $C_{l-1}$ and $C_{l-2}$ in (\ref{sdcl-def}) to $\bar{C}_{l-1}(\tau_l)$ and $\bar{C}_{l-2}(\tau_l)$ such that
\be
\bar{C}_{l-1}(\tau_l)=\bar{s}_l+\bar{d}_l(\tau_l),\qquad \bar{C}_{l-2}(\tau_l)=\bar{s}_l-\bar{d}_l(\tau_l).
\ee
With these, the anomalous dimension given by (\ref{cascsl}) is promoted to $\bar{\gamma}_l(\tau_l)$, where
\ba
\bar{\gamma}_l(\tau_l)&=&-\frac{1}{2}-\frac{1}{3}\bar{C}_{l-2}(\tau_l)\bar{C}_{l-1}(\tau_l)= -\frac{1}{2}-\frac{1}{3}(\bar{s}_l^2-\bar{d}_l(\tau_l)^2)\nn\\&=&
-\Bigl[\frac{1}{2}+\frac{1}{3}(\bar{s}_l^2-d_l^2)\Bigr]
+\Bigl[\frac{1}{3}(\bar{d}_l(\tau_l)^2-d_l^2)\Bigr].\label{gamma-gravity}
\ea
Thus the gravity side predicts that the anomalous mass dimensions of the mesons made out of the bifundamental chiral superfields are given by (\ref{gamma-gravity}). The terms in the first brackets in (\ref{gamma-gravity}) are constants and are exactly the values proposed in \cite{Hailu:2006uj} and given by (\ref{cascsl}) and (\ref{cldef}), with the first term of $-1/2$ being the value in Klebanov-Strassler and $-(\bar{s}_l^2-d_l^2)/3$ being the corrections. The terms in the second brackets in (\ref{gamma-gravity}), $(\bar{d}_l(\tau_l)^2-d_l^2)/3$, are nonlinear and depend on the scale of the theory (are functions of $\tau_l$).
Unfortunately, calculating the nonlinear corrections to the anomalous mass dimensions within the gauge in not an easy task and we could not compare the additional nonlinear corrections predicted by the supergravity solutions to gauge theory calculations.
The anomalous mass dimensions in the gauge theory are related to the physical running of the couplings and come from wavefunction renormalization involving nonholomorphic kinetic terms.

\subsection{D7-branes}

Let us recall that the $F_1$ flux we have at hand, given by (\ref{F1-tot1}), is not closed, $dF_1\ne 0$, for two reasons. One is simply because the magnitudes of $s$ and $d$ which give rise to the runnings of the dilaton and the axion change across duality transitions. Second, $dF_1\ne 0$ even for the flows between locations of duality transitions since $\tilde{\epsilon}_3$ is not closed. We will see by the end of our discussions in section \ref{sec-camd} that the sources are magnetic couplings of the axion to D7-branes filling 4-d spacetime and wrapping 4-cycles at the radial locations of duality transitions and invisible Dirac 8-branes which emanate from the D7-branes.
First let us rewrite the $F_1$ flux given by (\ref{F1-tot1}) or (\ref{F1-tot2}) in terms of the spacetime coordinates,
\be
F_1=\frac{s}{3+d}\tilde{\epsilon}_3=\frac{s}{3+d} \frac{v^{1/2}}{u^{1/4}}G^6=\frac{s}{3+d}(d\psi+\cos\theta_1d\phi_1+\cos\theta_2d\phi_2).
\label{F1-tot3}
\ee
Let us consider the flow between locations of duality transitions in which ${s}/{(3+d)}$ is a constant. Taking the Hodge star of (\ref{F1-tot3}),
\ba
\star F_1 &=& -\frac{s}{3+d}\, \frac{v^{1/2}}{h u^{1/4}} \,vol_4\wedge G^1\wedge G^2 \wedge G^3 \wedge G^4 \wedge G^5\nn\\
&=&-\frac{s}{3+d} \frac{u}{h}\sin \theta_1 \sin \theta_2 \,vol_4\wedge d\theta_1 \wedge d\phi_1 \wedge d\theta_2 \wedge d\phi_2\wedge d\tau,
\ea
where $vol_4=dx^0\wedge dx^1\wedge dx^2\wedge dx^3$ is the 4-d volume element.
Let us write \be F_1 = \star \mathcal{F}_9,\label{F1starF9}\ee where
\be
\mathcal{F}_9=\frac{s}{3+d} \frac{u}{h}\sin \theta_1 \sin \theta_2 \,vol_4\wedge d\tau\wedge d\theta_1 \wedge d\phi_1 \wedge d\theta_2 \wedge d\phi_2.\label{F9-defn}
\ee
But this $\mathcal{F}_9$ flux comes from D7-branes filling 4-d spacetime and wrapping the 4-cycle
\ba
\omega_4&=&\sin \theta_1 \sin \theta_2 \, d\theta_1 \wedge d\phi_1 \wedge d\theta_2 \wedge d\phi_2
\nn\\
&=&\frac{1}{u}G^1\wedge G^2\wedge G^3\wedge G^4.
\ea
Note also that $d\mathcal{F}_9=0$ and we write $\mathcal{F}_9=dC_8$, where the D7-branes potential is
\ba
C_8&=&\left(\int^\tau \frac{s}{3+d} \frac{u(\tau')}{h(\tau')}d\tau'\right)\sin \theta_1 \sin \theta_2 \,vol_4\wedge d\theta_1 \wedge d\phi_1 \wedge d\theta_2 \wedge d\phi_2\nn\\
&=& \left(\int^\tau \frac{s}{3+d} \frac{u(\tau')}{h(\tau')}d\tau'\right)\,\frac{1}{u(\tau)}vol_4\wedge G^1\wedge G^2\wedge G^3\wedge G^4.\label{C8-1}
\ea
Because the magnitudes of $s$ and $d$ change across each location of duality transition,
the D7-branes wrap $\omega_4$ at radial locations where Seiberg duality transitions take place.
In other words, a duality transition takes place as the theory flows across the D7-branes.
The values of the radial locations follow from the periodicity of $b_2$ in section \ref{sec-dm} and the expression for $h_1$ in section \ref{sec-solns}.

\subsection{Dirac 8-branes}

In this section we will see the magnetic coupling of the axion to the D7-branes via Dirac 8-branes whose worldvolume emanates from the worldvolume of the D7-branes. A general discussion of couplings between axion and 7-branes can be found in \cite{Bergshoeff:2007aa}. The relation between the Dirac 8-branes and the D7-branes is similar to the relation between a Dirac string and a magnetic monopole on which the string ends in 4-d.
First let us rewrite (\ref{F9-defn}) as
\ba
\mathcal{F}_9 &=&\frac{s}{3+d} \frac{1}{h} \,d\tau\wedge vol_4\wedge G^1\wedge G^2\wedge G^3\wedge G^4.\label{D9-defn}
\ea

\begin{figure}[htb]
\begin{center}
\leavevmode
\includegraphics[width=0.9\textwidth, angle=0]{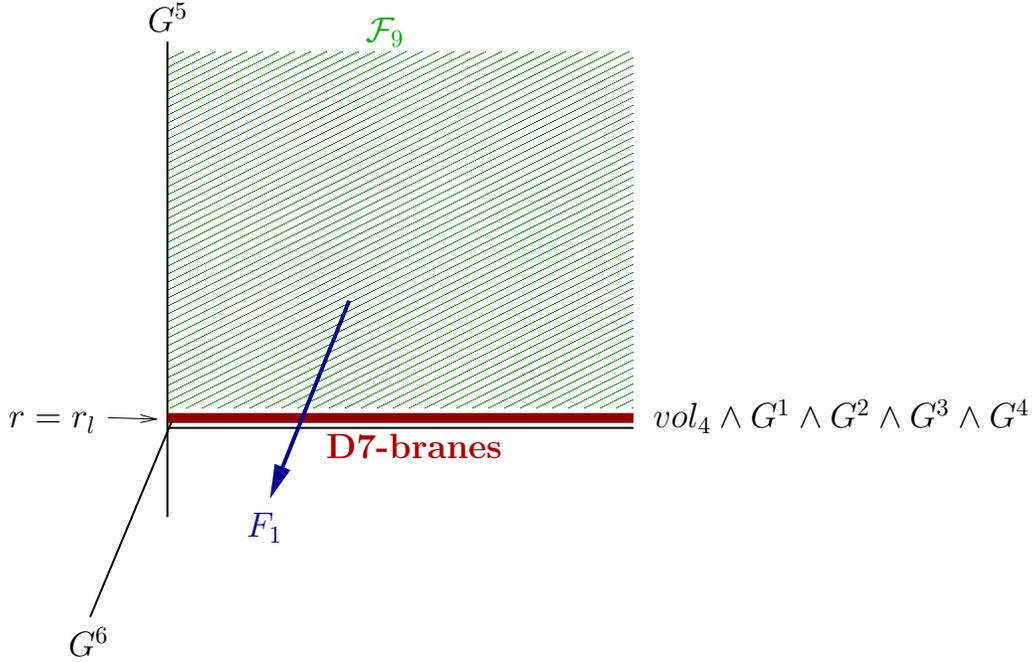}
\caption[]{D7-branes and $\mathcal{F}_9$ and $F_1$ fluxes. The figure schematically shows the worldvolume of D7-branes (the horizontal solid line) filling 4-d spacetime and wrapping 4-cycle at a radial location of duality transition $r=r_l$ in the extra space, the $\mathcal{F}_9=\bar{\mathcal{F}}_9-\star dC_0$ flux (the solid region) with the $\bar{\mathcal{F}}_9$ part carried by the worldvolume of Dirac 8-branes  which emanate from the D7-branes, and the 1-form $F_1$ flux (the vector pointing out of the plane).}
\label{fig-d7d8}
\end{center}
\end{figure}

The magnetic 2-form current $\star \mathcal{J}_8$ of the D7-branes, again in the flow between duality transitions, is
\ba
\star \mathcal{J}_8&=&d \star \mathcal{F}_9=dF_1=-\frac{s}{3+d}(\sin\theta_1 d\theta_1\wedge d\phi_1+\sin\theta_2 d\theta_2\wedge d\phi_2)\nn\\
&=& \frac{s}{3+d}\frac{1}{\sqrt{u}}(G^1\wedge G^2+G^3\wedge G^4),\label{JD8-1}
\ea
where we have used the solutions in section \ref{sec-solns} in relating the expressions in the two lines.
The point is that $\star \mathcal{J}_8$ is closed and comes from the terms in  $F_1$ which make $dF_1\ne 0$. The D7-branes carry magnetic charge and their electric duals are D(-1)-branes which are point-like both in space and in time.
Note also that it follows from (\ref{C8-1}) and (\ref{JD8-1}) that
\ba
C_8\wedge \star \mathcal{J}_8=0.\label{C8stJ8-1}
\ea
We will see in the next section that this vanishing coupling of $\star \mathcal{J}_8$ to the D7-branes potential $C_8$ gives the runnings of the axion and the dilaton.
Let us separate out the terms which make $dF_1\ne 0$ and rewrite $F_1$ as
\be
F_1-\star \bar{\mathcal{F}}_9=dC_0,\label{F1-tot4}
\ee
where
\ba
C_0&=&\frac{s}{3+d}\psi\label{C0-1}\ea  is the axion which describes the magnetic D7-branes charges, $dC_0$ is constant for flows between duality transitions, and
\ba \star  \bar{\mathcal{F}}_9&=&\frac{s}{3+d}(\cos\theta_1d\phi_1+\cos\theta_2d\phi_2),\label{stD9b}
\ea
contains the terms which make $dF_1\ne0$.
The magnetic flux $\mathcal{F}_9=dC_8=-\star dC_0+\bar{\mathcal{F}}_9$ is a sum of the flux from the magnetic dual to the axion, $-\star dC_0$, and the flux $\bar{\mathcal{F}}_9$.
We see that we can think of $\bar{\mathcal{F}}_9$ as the worldvolume a hypothetical hypersurface of Dirac 8-branes with 8 spacial dimensions coming out of the worldvolume of the D7-branes. The configurations of the D7-branes and the $\mathcal{F}_9$ and the $F_1$ fluxes for the supergravity system we have here are schematically shown in figure \ref{fig-d7d8}.
Note that $\star \mathcal{J}_8=d \star \mathcal{F}_9=d \star \bar{\mathcal{F}}_9$ and these quantities are related to the corrections to the anomalous mass dimension,parameterized by $s$ and $d$, on the gauge theory side. The D7-branes charge at a radial location of duality transition is $\int_0^{4\pi}dC_0=4\pi s/(3+d)$ and demanding that the charge be integral imposes a constraint. For the values of $s_l$ and $d_l$ given by (\ref{sdcl-def}) and (\ref{cldef}) with $K=N/M$, this requires that $4M/(2K+3-2l)$ be integers which is accommodated for appropriate choices of $M$ and $K$.

\subsection{Magnetic coupling of axion to D7-branes via invisible Dirac 8-branes\label{sec-camd}}

Now we want to write the supergravity action which includes the magnetic coupling of the axion to the D7-branes via the Dirac 8-branes and show that the relations among $F_1$, $\mathcal{F}_9$, $\mathcal{F}_8$, and $\mathcal{J}_8$ we have found follow form the equations of motion. Moreover, nonzero $F_1$ flux amounts to a running dilaton because of the relation in (\ref{dAF1isd}).
First let us summarize the relations we have in (\ref{F1-tot3}), (\ref{JD8-1}) and (\ref{stD9b}),
\be dF_1=d\star \mathcal{F}_9=d\star\bar{\mathcal{F}}_9=\star\mathcal{J}_8.\label{dF1D8J8}\ee
\comment{We can define an effective D7-charge $q_7$ in terms of number of D7-branes $N_7$ using the expression for $C_0$ given by (\ref{C0-1}) as
\be
q_7=\int dC_0= \frac{s}{3+d}N_7 \int_0^{4\pi} d\psi\ee
and a single unit of D7-brane then follows as
\be
N_7=\frac{3+d}{4\pi s}\int dC_0=1.\ee}

The type IIB bosonic action in string frame is \cite{Polchinski:1998rr}
\begin{eqnarray}
S_{10}&=&\frac{1}{2\kappa^2}\int \Bigl( d^{10}x
\sqrt{-G}\Bigl[\frac{1}{g_s^2}e^{-2\Phi}\left(R+4(\partial \Phi)^2-\frac{1}{12}g_sH_3^2\right)-\frac{1}{2}
F_1^2  \nonumber \\
&&-\frac{1}{12}g_s
{F}_3^2-\frac{1}{4.5!}g_s^2{F}_5^2\Bigr]
-\frac{1}{2}g_s^2C_4\wedge {F}_3 \wedge {H}_3\Bigl), \label{s2bsgra-IIB}
\end{eqnarray}
where
$G$ is the determinant of the metric, $R$ is
the Ricci scalar, and $\kappa$ is the gravitational constant in ten dimensions. Now as in the discussions in \cite{Bergshoeff:2007aa} let us add the following terms to the action,
\begin{eqnarray}
\Delta S_{10}&=&\frac{1}{2\kappa^2}\Bigl(\int  d^{10}x
\Bigl[\frac{1}{8!}\epsilon^{M M_1\cdots M_9}(F_1)_M \partial_{M_1}(C_8)_{M_2 \cdots M_9}\Bigl]\nonumber
-\int {C}_8\wedge \star\mathcal{J}_8\Bigr)\nn\\&&
-\int_{\mathcal{M}^8}d^8 \sigma \, {\frac{1}{g_s}e^{-\Phi}}T_7\sqrt{-\det G_{ab}}, \label{s2bsgra-IIB-D7}
\end{eqnarray}
where the values of $F_1$, $C_8$, and $\star\mathcal{J}_8$ are given respectively by (\ref{F1-tot3}), (\ref{C8-1}), and (\ref{JD8-1}).
The first term is a magnetic coupling of the axion to the D7-branes. The second term in (\ref{C8stJ8-1}) is the coupling of the D7-branes to the current of Dirac 8-branes and it is an interesting piece, it is exactly zero as we saw in (\ref{C8stJ8-1}). The third term is the worldvolume of the D7-branes with coordinates $\sigma^a$ (on the branes) the metric (with the indices $a$ and $b$) and the dilaton pulled back to the worldvolume $\mathcal{M}^8$ of the D7-branes and $T_7$ is the tension.

Varying $S_{10}+\Delta S_{10}$ with respect to $F_1$ gives exactly (\ref{F1starF9}) and that with respect to $C_8$ gives  exactly (\ref{dF1D8J8}). Note that these two equations are not independent.
But we can rewrite the second term in (\ref{s2bsgra-IIB-D7}) with (\ref{dF1D8J8}) as
\ba
-\int {C}_8\wedge \star\mathcal{J}_8&=&-\int {C}_8\wedge dF_1= \int d{C}_8\wedge F_1-\int d({C}_8\wedge F_1)\nn\\&=&-\int F_1\wedge d{C}_8 +\,\mathrm{surface\, term}.\label{C8J8F1}
\ea
With (\ref{C8J8F1}), $\Delta S_{10}$ given by (\ref{s2bsgra-IIB-D7}) reduces (up to surface terms) to
\begin{eqnarray}
\Delta S_{10}&=&-\sum_{l=1}^{K-1}\int_{\mathcal{M}^8}d^8 \sigma \, {\frac{1}{g_s}e^{-\Phi(r_l)}}T_7(r_l)\sqrt{-\det G_{ab}(r_l)}, \label{s2bsgra-IIB-D7-red}
\end{eqnarray}
where we have explicitly put in the radial locations $r_l$ of duality transitions where the D7-branes fill 4-d spacetime and wrap the 4-cycle $\omega_4$ and summed over all $K-1$ locations.
The first two terms in $\Delta S_{10}$ given by (\ref{s2bsgra-IIB-D7}) which give the couplings of the axion to the D7-branes via the Dirac 8-branes vanish (up to surface terms) and the magnetic $C_8$ potential has completely disappeared from (\ref{s2bsgra-IIB-D7-red}) and the action with its effects seen only in the runnings of the dilaton and the axion.
What is happening is that the components of the $F_1$ flux which make $dF_1\ne 0$ are communicated from the D7-branes via the invisible Dirac 8-branes whose worldvolume emanates from the world volume of the magnetic D7-branes in the same way a Dirac string emanates from a magnetic monopole. The remaining additional term (\ref{s2bsgra-IIB-D7-red}) in the action, in the end, is simply the worldvolume of the D7-branes which fill up 4-d spacetime and wrap the $\omega_4$ cycles at the locations of duality transitions. Because the duality transitions occur as the theory flows across the D7-branes, the D7-branes also serve as gravitational source for Seiberg duality transitions.

\section{Conclusions and discussions}

We have presented a gauge/gravity duality construction with explicit supergravity solutions to a cascading $\mathcal{N}=1$ supersymmetric $SU(N+M)\times SU(N)$ gauge theory with bifundamental chiral superfields and a quartic tree level superpotential which includes the corrections proposed in \cite{Hailu:2006uj} to the anomalous mass dimension in the Klebanov-Strassler throat. Including the corrections leads to an amazingly consistent picture with  the mass dimension of the operator in the quartic tree level superpotential falling just within a narrow window in which the operator stays relevant with the mesons behaving as scalars and the duality cascade ending in $SU(2M)\times SU(M)$ gauge theory as we expect and, at the same time, revealing novel structures on the gravity side including providing gravitational source for Seiberg duality transitions and features which could be useful for constructing cosmological models which allow to study possibilities for probing distinct stringy signatures from the early universe.

It is worth emphasizing that our construction is the first explicit example of a gauge/gravity duality mapping with a running dilaton or a running axion which is an important component towards finding gravity theories which are dual to gauge theories with physically more interesting renormalization group flows such as pure confining gauge theories.
The picture which emerges out of the solutions on the gravity side is quite interesting. We showed that magnetic couplings of the axion to D7-branes filling 4-d spacetime and wrapping 4-cycles and invisible Dirac 8-branes which emanate from the D7-branes are the sources for the runnings of the dilaton and the axion. We saw that all these consistently follow as solutions to the equations of motion of the supergravity action. The duality transitions occur as the theory flows across locations of $AdS_5$ radii where the D7-branes fill up 4-d spacetime and wrap 4-cycles and, therefore, the D7-branes also serve as gravitational source for Seiberg duality transformations.

With the supergravity solutions in hand, we went backwards and found the anomalous mass dimensions of the meson operators in the gauge theory that the gravity theory predicts whose constant pieces reproduced the same values as the ones proposed in \cite{Hailu:2006uj} and additional nonlinear corrections which depend on the scale of the theory as we expect.
The supergravity solutions might be useful for studying the nonperturbative dynamics of the dual gauge theory and the gravity description might reveal additional features. The supergravity flow in the tip region of the throat involves $SU(2M)$ gauge theory with $2M$ flavors.

Similar techniques could be useful for studying and finding gravity flows which correspond to interesting gauge theories with $\mathcal{N}=1$ supersymmetry such as pure confining gauge theories in four dimensions. One may, for instance, start with the renormalization group flow of a gauge theory, study its implications to the runnings of the dilaton and the fluxes which with the equations for type IIB flows with $\mathcal{N}=1$ supersymmetry given in \cite{Hailu:2007ae} and appropriate metric and flux ansatz might allow to construct a dual gravity theory or at least guide towards finding it.

Currently ongoing and future observations in cosmology are (and will be) providing high precision data from the early universe at energy scales which provide a natural laboratory for stringy physics. One of the issues in string cosmology scenarios is a difficulty in constructing models which could allow to probe uniquely stringy signatures.  The gravity theory we have here involves novel features which could be useful for constructing cosmological models which might possibly allow to probe distinct stringy signatures from the early universe as discussed in \cite{Hailu:2006uj}. Two key features are first, the steps in the warp factor due to the change in the magnitudes of $s_l$ and $d_l$ across duality transitions and second, the radial locations of the steps are correlated. Such features might allow to construct models which mighty possibly allow to read or predict stringy signatures in the anisotropies in the cosmic microwave background.

A gravity dual to pure confining $\mathcal{N}=1$ supersymmetric $SU(N)$ gauge theory which has correct gauge coupling running and appropriate pattern of chiral symmetry breaking is constructed in \cite{Hailu:pcgt} using results in this article.

\section*{Acknowledgements}

We are grateful to Henry Tye for helpful discussions.
This research is supported in part by the National Science Foundation under grant number
NSF-PHY/03-55005.

\appendix

\section{$SU(3)$ components of the torsion and the fluxes\label{app-dec}}

The components in the decompositions of the torsion and the fluxes in $SU(3)$ representations in section \ref{sec-ftc} for the ansatz in section \ref{sec-ansatz} and used in finding the solutions in section \ref{sec-solns} are given below.

\begin{eqnarray}\label{W36bZ}
W_3^{(6)}&=&\frac{1}{8} i e^{-g-3 p-\frac{3 x}{2}} \nonumber \\ &&(({Z^1}{
   {\wedge} }{\bar{Z}^{\bar{1}}}{ {\wedge} }{\bar{Z}^{\bar{3}}}-{Z^2}{ {\wedge}
   }{\bar{Z}^{\bar{2}}}{ {\wedge} }{\bar{Z}^{\bar{3}}}) ((a^2+e^{2 g}-1)
   \mathcal{A}-2 e^{6 p+2 x} (\mathcal{A} \mathcal{B} a'+e^{g} (\mathcal{B}^2-1)
   g'))\nonumber \\ &&+\mathcal{B} ({Z^1}{ {\wedge}
   }{\bar{Z}^{\bar{2}}}{ {\wedge} }{\bar{Z}^{\bar{3}}}-{\bar{Z}^{\bar{1}}}{
   {\wedge} }{Z^2}{ {\wedge} }{\bar{Z}^{\bar{3}}}) (a^2+e^{2 g}-1-2 e^{6 p+2
   x} (\mathcal{B} a'-e^{g} \mathcal{A} g'))
   \nonumber \\ &&
   +({\bar{Z}^{\bar{1}}}{ {\wedge}
   }{\bar{Z}^{\bar{2}}}{ {\wedge} }{Z^3}) ((a^2-e^{2 g} -1)\mathcal{B} -2 e^{g} \mathcal{A}
   a+2 e^{6 p+2 x} (a \mathcal{A}\nonumber \\ &&-a'+e^{g} (\mathcal{B}
   (\mathcal{A}'+1)-\mathcal{A} \mathcal{B}')))
   )
\end{eqnarray}
\begin{eqnarray}\label{W43bZ}
W_4^{(\bar{3})}&=&
\frac{1}{4} e^{-g-3 p-\frac{3 x}{2}} {\bar{Z}^{\bar{3}}} \Bigl((a^2-e^{2
   g}-1) \mathcal{A}+2 e^{g} (a \mathcal{B}+e^{6 p+2 x} x')\Bigr)
\end{eqnarray}
\begin{eqnarray}\label{W53bZ}
W_5^{(\bar{3})}&=&
-\frac{1}{4} e^{-g+3 p+\frac{x}{2}} {\bar{Z}^{\bar{3}}} (e^{g} (2 \mathcal{A}+6
   p'-x')-2 a \mathcal{B})
\end{eqnarray}
\begin{eqnarray}\label{W11bZ}
W_1^{(1)}&=&
-\frac{1}{6} e^{-g-3 p-\frac{3 x}{2}} \Bigl(-\mathcal{B} a^2+2 e^{g} \mathcal{A} a+e^{2
   g} \mathcal{B}+\mathcal{B}+2 e^{6 p+2 x} (a \mathcal{A}+a'\nonumber \\ &&+e^{g} (-\mathcal{A}'
   \mathcal{B}+\mathcal{B}+\mathcal{A} \mathcal{B}'))\Bigr)
\end{eqnarray}
\begin{eqnarray}\label{W2Z}
W_2^{(8)}&=&
\frac{2}{3} i e^{-g-3 p-\frac{3 x}{2}} (-\frac{1}{2} ({Z^3}{
   {\wedge} }{\bar{Z}^{\bar{3}}}) (\mathcal{B} a^2-2 e^{g} \mathcal{A} a-e^{2 g} \mathcal{B}-\mathcal{B}+e^{6
   p+2 x} (a \mathcal{A}+a'\nonumber \\ &&+e^{g} (-\mathcal{A}' \mathcal{B}+\mathcal{B}+\mathcal{A}
   \mathcal{B}')))\nonumber \\ &&-\frac{3}{4} e^{6 p+2 x} \mathcal{B} ({Z^1}{
   {\wedge} }{\bar{Z}^{\bar{2}}}-{\bar{Z}^{\bar{1}}}{ {\wedge} }{Z^2}) (a+\mathcal{A}
   a'+e^{g} \mathcal{B} g')\nonumber \\ &&-\frac{3}{4} ({Z^2}{ {\wedge} }{\bar{Z}^{\bar{2}}})
   (-\mathcal{B} a^2+2 e^{g} \mathcal{A} a+(1+e^{2 g}) \mathcal{B}+e^{6 p+2 x}
   (e^{g} (\mathcal{B} (\mathcal{A}'+\mathcal{A} g'-1)\nonumber \\ &&-\mathcal{A} \mathcal{B}')-\mathcal{B}^2
   a'))\nonumber \\ &&-\frac{1}{4} ({Z^1}{ {\wedge} }{\bar{Z}^{\bar{1}}}) (-\mathcal{B}
   a^2+2 e^{g} \mathcal{A} a+e^{2 g} \mathcal{B}+\mathcal{B}+e^{6 p+2 x} (2 a
   \mathcal{A}-(3 \mathcal{B}^2-2) a'\nonumber \\ &&+e^{g} (\mathcal{B} (\mathcal{A}'+3 \mathcal{A}
   g'-1)-\mathcal{A} \mathcal{B}'))))
\end{eqnarray}
\begin{eqnarray}\label{H36bZ}
H_3^{(6)}&=&
\frac{1}{8} i e^{-g+3 p-\frac{x}{2}} (({Z^1}{ {\wedge} }{\bar{Z}^{\bar{1}}}{ {\wedge}
   }{\bar{Z}^{\bar{3}}}-{Z^2}{ {\wedge} }{\bar{Z}^{\bar{2}}}{ {\wedge} }{\bar{Z}^{\bar{3}}}) ((a^2+e^{2
   g}+1) \mathcal{A} {h_1}'+2 a \mathcal{A} {h_2}')\nonumber \\ &&+({Z^1}{ {\wedge} }{\bar{Z}^{\bar{2}}}{ {\wedge} }{\bar{Z}^{\bar{3}}}-{\bar{Z}^{\bar{1}}}{ {\wedge}
   }{Z^2}{ {\wedge} }{\bar{Z}^{\bar{3}}}) ((a^2+e^{2 g}+1) \mathcal{B} {h_1}'+2 a \mathcal{B}
   {h_2}')\nonumber \\ &&+({\bar{Z}^{\bar{1}}}{ {\wedge} }{\bar{Z}^{\bar{2}}}{
   {\wedge} }{Z^3}) (2 e^{g} {h_2}+(\mathcal{B} a^2-2 e^{g} \mathcal{A} a-2 e^{g} \mathcal{B} \sinh
   (g)) {h_1}'\nonumber \\ &&+(2 a \mathcal{B}-2 e^{g} \mathcal{A}) {h_2}'))
\end{eqnarray}
\begin{eqnarray}\label{H33bZ}
H_3^{(\bar{3})}&=&
\frac{1}{4} e^{-g+3 p-\frac{x}{2}} {\bar{Z}^{\bar{3}}} (((a^2-e^{2 g}+1) \mathcal{A}+2 e^{g} a
   \mathcal{B}) {h_1}'+(2 a \mathcal{A}+2 e^{g} \mathcal{B}) {h_2}')
\end{eqnarray}
\begin{eqnarray}\label{H31Z}
H_3^{(1)}&=&
\frac{1}{6} e^{-g+3 p-\frac{x}{2}} (-2 e^{g} {h_2}+(\mathcal{B} a^2-2 e^{g} \mathcal{A}
   a-2 e^{g} \mathcal{B} \sinh g) h'_1\nonumber \\ &&+(2 a \mathcal{B}-2 e^{g} \mathcal{A})
   h'_2)
\end{eqnarray}
\begin{eqnarray}\label{F36bZ}
F_3^{(6)}&=&
\frac{1}{8} e^{-g+3 p-\frac{x}{2}-\Phi} P (-\mathcal{A}(({Z^1}{ {\wedge} }{\bar{Z}^{\bar{1}}}{ {\wedge} }{\bar{Z}^{\bar{3}}})-({Z^2}{ {\wedge} }{\bar{Z}^{\bar{2}}}{
   {\wedge} }{\bar{Z}^{\bar{3}}}))(a^2-2 b a+e^{2 g}+1)
   \nonumber \\ && -\mathcal{B}
   (({Z^1}{ {\wedge} }{\bar{Z}^{\bar{2}}}{ {\wedge} }{\bar{Z}^{\bar{3}}})-
   ({\bar{Z}^{\bar{1}}}{ {\wedge} }{Z^2}{ {\wedge} }{\bar{Z}^{\bar{3}}}))(a^2-2 b a+e^{2 g}+1)
   \nonumber \\ &&-({\bar{Z}^{\bar{1}}}{ {\wedge} }{\bar{Z}^{\bar{2}}}{ {\wedge}
   }{Z^3}) (e^{2 g} \mathcal{B}-(a^2-2 b a+1) \mathcal{B}+2 e^{g} (\mathcal{A}
   (a-b)+b')))
\end{eqnarray}
\begin{eqnarray}\label{F33}
F_3^{(\bar{3})}&=&
-\frac{1}{4} i e^{-g+3 p-\frac{x}{2}-\Phi} P {\bar{Z}^{\bar{3}}} (\mathcal{A} (-a^2+2 b
   a+e^{2 g}-1)-2 e^{g} (a-b) \mathcal{B})
\end{eqnarray}
\begin{eqnarray}\label{F31}
F_3^{(1)}&=&
\frac{1}{6} i e^{-g+3 p-\frac{x}{2}-\Phi} P (-2 e^{g} \mathcal{A} (a-b)+(a
   (a-2 b)-e^{2 g}+1) \mathcal{B}+2 e^{g} b')
\end{eqnarray}
\begin{eqnarray}\label{F53b}
F_5^{(\bar{3})}&=&
\frac{1}{16} i e^{-g+3 p-\frac{3x}{2}}{\bar{Z}^{\bar{3}}} K (a\mathcal{A}\mathcal{B}+e^{g}(\mathcal{B}^2-1))
\end{eqnarray}
\begin{eqnarray}\label{F13b}
F_{1}^{(\bar{3})}&=&\frac{i}{2}\frac{s}{3+d} e^{3p+\frac{x}{2}}\bar{Z}^{\bar{3}}
\end{eqnarray}

\section{The Klebanov-Strassler solution}

Here we show that the solutions we have written in section \ref{sec-solns} contain the Klebanov-Strassler solution \cite{Klebanov:2000hb} for $s=d=0$. First, note that in this case the anomalous mass dimension given by (\ref{cascsl}) reduces to $-1/2$, the dilaton given by (\ref{phi-soln}) is constant and is set to zero, and the $F_1$ flux given by (\ref{F1-tot1}) vanishes,
\be\gamma=-1/2,\quad \Phi=0,\quad F_1=0.\ee
Moreover, the $\tau$ defined in (\ref{taut-not}) reduces to
\be\tau=3t=3\ln(r/r_c).\label{taut-ks}\ee
The expressions for $b$, $h_1$, $h_2$, $a$, $g$, $A$, $B$, and $K$ are the same as given by (\ref{b-sol})-(\ref{Ksolna}) with $\tau$ given by (\ref{taut-ks}) and $\Phi=0$.
The equations for $v$, $u$ and $h$ given by (\ref{ve-1}), (\ref{uh-sola}) and (\ref{h-eqa}) reduce, with $s=d=0$, to
\be\label{ve-1-ks}
v'+2\coth \tau\, v-3=0,
\ee
\be
\frac{d}{d\tau}\left(\frac{u}{h}\right)=\frac{2}{v},\label{uh-sola-ks}
\ee
\be
h'+ \frac{1}{4}g_s K\frac{h}{u}=0\label{h-eqa-ks}
\ee
Using (\ref{h2-soln}) with $\Phi=0$ in (\ref{Ksolna}),
\be
K=\frac{1}{8} \alpha'^2 g_sM^2 \left(\tau \coth \tau-1\right) \left(\tau {\csch}^2 \tau-\coth
   \tau\right).\label{Ksolna-ks}
\ee
Solving (\ref{ve-1-ks}) with $v(0)=0$ gives the solution for $v$,
\be
v=\frac{3}{2} \left(\coth \tau-\tau\, {\csch}^2 \tau\right).\label{ve-2-ks}
\ee
Using (\ref{ve-2-ks}) in (\ref{uh-sola-ks}) gives
\be
u=(\sinh (2 \tau)-2 \tau)^{2/3} h.\label{u-soln-ks}
\ee
Finally, using (\ref{Ksolna-ks}) and (\ref{u-soln-ks}) for $h/u$ in (\ref{h-eqa-ks}),
\be
h'=\frac{1}{64} \alpha'^2 g_s^2M^2 (\tau \coth \tau-1)  \left(\sinh (2
   \tau)-2 \tau \right)^{1/3} {\csch}^2 \tau.
\ee
Thus our solutions reproduce the Klebanov-Strassler solution in the forms written in \cite{Butti:2004pk} as a special case for $s=d=0$ up to different choices of some overall constants.


\bibliographystyle{JHEP}


\providecommand{\href}[2]{#2}\begingroup\raggedright\endgroup

\end{document}